\documentclass[aps,prl,twocolumn,showpacs,superscriptaddress]{revtex4} 
\usepackage{graphicx}
\usepackage{bm}
\usepackage[hidelinks]{hyperref}
\usepackage{dcolumn,xcolor,ulem}
\usepackage{amsmath} 
\usepackage{amssymb}
\usepackage{amsfonts}
\usepackage[latin1]{inputenc}

\begin{document}

\title{Grains unchained: local fluidization of a granular packing by focused ultrasound}

\author{Pierre~Lidon} \affiliation{Universit\'{e} de Lyon, Laboratoire de Physique, \'Ecole Normale Sup\'{e}rieure de Lyon, CNRS UMR 5672, 46 all\'{e}e d'Italie, 69364 Lyon Cedex 07, France}
\author{Nicolas~Taberlet}
 \affiliation{Universit\'{e} de Lyon, Laboratoire de Physique, \'Ecole Normale Sup\'{e}rieure de Lyon, CNRS UMR 5672, 46 all\'{e}e d'Italie, 69364 Lyon Cedex 07, France}
 \affiliation{Universit\'e de Lyon, UFR de Physique, Universit\'e Claude Bernard Lyon I, Lyon, France}
\author{S\'{e}bastien~Manneville} \affiliation{Universit\'{e} de Lyon, Laboratoire de Physique, \'Ecole Normale Sup\'{e}rieure de Lyon, CNRS UMR 5672, 46 all\'{e}e d'Italie, 69364 Lyon Cedex 07, France}

\date{\today}

\begin{abstract}
We report experimental results on the dynamics of a granular packing submitted to high-intensity focused ultrasound. Acoustic radiation pressure is shown to remotely induce local rearrangements within a pile as well as global motion around the focal spot in an initially jammed system. We demonstrate that this fluidization process is intermittent for a range of acoustic pressures and hysteretic when the pressure is cycled. Such a first-order-like unjamming transition is reproduced in numerical simulations in which the acoustic pressure field is modeled by a localized external force. Further analysis of the simulated packings suggests that in the intermittent regime unjamming is not associated with any noticeable prior structural signature. A simple two-state model based on effective temperatures is proposed to account for these findings.  
\end{abstract}

\maketitle

\section{Introduction} 

Besides noninvasive imaging at low intensities, ultrasound is commonly used at high acoustic powers in therapeutic applications such as kidney stone lithotripsy or tumor treatment \cite{Kennedy:2005,terHaar:2007}. Indeed high-intensity focused ultrasound (HIFU) can induce strong mechanical stresses and shear wave propagation in biological tissues \cite{Bercoff:2004b} or even cell membrane permeabilization and rupture through cavitation \cite{Marmottant:2003,Prentice:2005}. Surprisingly, HIFU has not yet been used in the fundamental context of soft disordered ``jammed'' systems although the physics of jamming has attracted tremendous attention over the last two decades \cite{Cates:1998,Liu:1998,Olsson:2007,Liu:2010,Ikeda:2012,Lerner:2012a}. Soft jammed materials range from emulsions, foams and colloidal suspensions to granular materials and typically show a transition from fluidlike to solidlike behavior upon increasing the volume fraction \cite{Howell:1999,Silbert:2005,Berthier:2005,Keys:2007,Jaeger:2015}. Among them granular packings have emerged as model amorphous systems to study such a jamming transition at zero temperature \cite{Liu:2010,Brujic:2005,Dauchot:2005,Majmudar:2007,Behringer:2008,Bi:2011,Otsuki:2011,Somfai:2007,Bandi:2013}.

To date previous works probing granular packings with intense acoustic waves have been dedicated to dry grains under unfocused, low-frequency ultrasound, reporting rather limited fluidization effects such as creep motion of weakened contacts \cite{Jia:2011,Espindola:2012} or shock wave propagation \cite{Gomez:2012,vandenWildenberg:2013a,vandenWildenberg:2013b}. In this article we introduce HIFU as a useful, versatile tool to remotely exert a {\it localized} volumic force onto a dense three-dimensional granular assembly and to trigger its unjamming. Above a critical ultrasonic intensity, stresses due to the so-called acoustic radiation pressure \cite{Yosioka:1955,Chen:1996} become high enough to induce large-scale flows within the pile. We show both experimentally and numerically through space- and time-resolved velocity measurements that this local unjamming transition displays intermittency and hysteresis. Further analysis of the simulated stress tensor reveals that in the intermittent regime, after an acoustic pulse has unjammed the system, the state of the packing at rest is undistinguishable from the previously jammed state so that unjamming cannot be predicted from standard static observables. A simple two-state model involving an effective noise temperature reproduces qualitatively the observed features.

We first describe our experimental setup and numerical methods in Sect.~\ref{sec:matmeth}. There we also emphasize the structure of the ultrasonic field and show that acoustic radiation pressure dominates over other nonlinear acoustic effects. Experimental and numerical results are detailed in Sect.~\ref{sec:results}. Finally, in Sect.~\ref{sec:discuss}, we analyze the structure of the simulated packings, devise the two-state model and compare our work to previous studies on more classical fluidization modes in granular materials. We conclude this article by listing a number of questions that remain unanswered and some possibilities for future research.

\section{Materials and methods}
\label{sec:matmeth}

\subsection{Experimental setup}

\begin{figure*}[!tb]
\centering	
\includegraphics[scale=0.95]{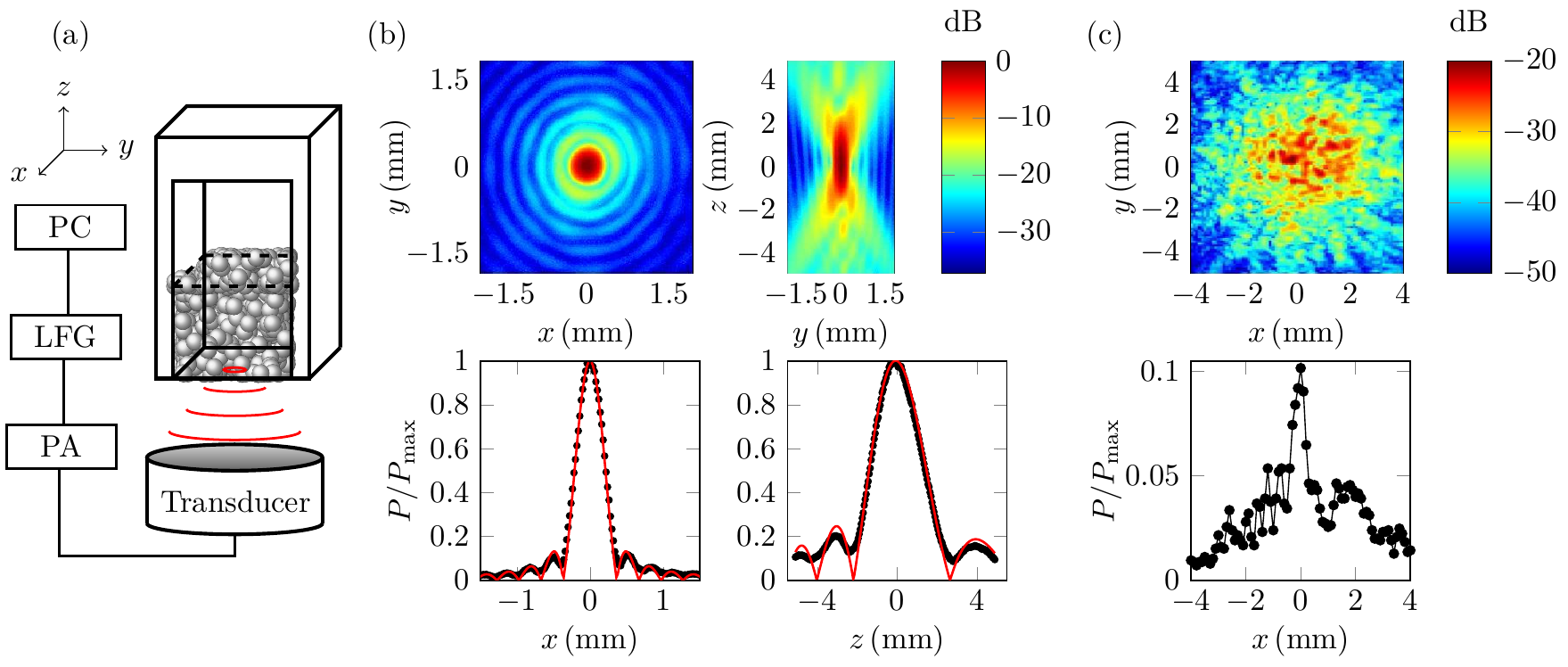}
\caption{(a) Schematics of the experimental setup. The pressure field emitted by the transducer is focused around the thin membrane that constitutes the cell bottom and about 1~mm behind the wall of the cell facing the CCD camera (not shown). ``PC'' stands for computer, ``LFG'' for low frequency generator and ``PA'' for power amplifier.
(b)~Pressure measurements in the absence of the experimental cell: (top) $P(x,y,0)$ across the horizontal focal plane at $z=0$ and $P(0,y,z)$ along the vertical plane at $x=0$ and (bottom) $P(x,0,0)$ and $P(0,0,z)$ normalized by the pressure $P_{\rm max}$ measured at the focal point. The red solid lines in the bottom left and right panels correspond respectively to Eqs.~(\ref{eq:Px}) and (\ref{eq:Pz}) with no adjustable parameter.
(c)~Pressure measurements in the presence of grains: (top) $P(x,y,z_0)$ across a horizontal plane located at $z_0\simeq 1$~mm, just above a couple of grain layers and (bottom) $P(x,0,z_0)$ normalized by the pressure $P_{\rm max}$ measured at the focal point in the absence of grains. The color scales correspond to the acoustic intensity in decibels relative to $P_{\rm max}$, i.e. to $20\log(P/P_{\rm max})$.}
\label{fig:setup}
\end{figure*}

As sketched in Fig.~\ref{fig:setup}(a) a dense packing of spherical beads is submitted to mechanical solicitations from HIFU generated by a high-power hemispherical piezoelectric transducer (Imasonic, diameter $25 \, \text{mm}$, center frequency $5 \, \text{MHz}$, wavelength $300 \, \mu\text{m}$ in water at room temperature) whose acoustic impedance is matched with water.
Polydisperse glass beads (USF Matrasur, batch F03 7911, density $2.5 \, \text{g.cm}^{-3}$) were sieved to achieve a mean diameter $d=550 \, \mu\text{m}$ with a standard deviation of $80 \, \mu\text{m}$.
These grains are first poured up to a height of roughly $1\,\text{cm}$ into a cell of size $0.5 \, \text{cm} \times 1 \, \text{cm} \times 3 \, \text{cm}$ filled with distilled water and whose open face is closed with a thin plastic membrane that has been checked to be transparent to ultrasound. The cell is finally reversed upside down so that the grains sediment under gravity and rest onto the membrane. Both the cell and the ultrasonic transducer are immersed in a large water tank ($\sim 1\,\text{L}$) filled with water at $20.0 \pm 0.5\,^\circ \text{C}$.

Images of the granular packing are captured with a CCD camera (Baumer HXC20) at up to 300~fps. Bursts of sinusoidal waves at $5 \, \text{MHz}$ and of duration $0.5 \, \text{s}$, hereafter referred to as ``pulses'',  are emitted every $2 \, \text{s}$ using a low frequency generator (Agilent 33522A) and a power amplifier (Kalmus 150C). The duty cycle is limited to $25\%$ to avoid heating of the transducer and the maximum achievable acoustic pressure at focus is $1.2 \, \text{MPa}$. In all cases the rest time of 1.5~s between two pulses is long enough so that all grains in the packing come to a complete stop. No cavitation bubbles were detected during or between pulses in both the water tank and the experimental cell. Prior to any experiment, the pile is submitted to two successive cycles of ultrasonic pulses whose amplitude is progressively increased up to the maximum acoustic power and then decreased back to 0, in order to reach a reproducible initial state (see also the discussion on hysteresis in Sect.~\ref{sec:hysteresis}). Next various protocols and analyses are applied: time-resolved studies of grain motion within a single pulse (Sect.~\ref{sec:unjamming}), long series of successive pulses with a constant amplitude (Sect.~\ref{sec:intermittency}), and cycles of pulses with increasing then decreasing amplitudes (Sect.~\ref{sec:hysteresis}). We checked that all our results are qualitatively unchanged when the pulse duration is varied. 

\subsection{Pressure field characterization} 

Figure~\ref{fig:setup}(b) shows pressure field measurements $P(x,y,z)$ obtained by scanning a needle hydrophone (Precision Acoustics 1659SN with active element diameter 40~$\mu$m) in water in the absence of the experimental cell. For these measurements we use short, low-intensity pulses of sinusoidal waves at frequency $f=5$~MHz with a fixed amplitude, a duration of 10~$\mu$s and a pulse repetition frequency of 1~kHz. The output voltage from the hydrophone is digitized by an oscilloscope (Agilent 33522A) and the pressure amplitude $P(x,y,z)$ is computed from an average over 1,000 successive pulses. The position of the hydrophone is controlled by three-axis precision translation stages (Physik Instruments M-410PD) and the origin $(x=0,y=0,z=0)$ is taken at the focal point where the pressure amplitude reaches its maximum value $P_{\rm max}$.

The two-dimensional scans of Fig.~\ref{fig:setup}(b), performed with steps $\delta x=\delta y=20$~$\mu$m and $\delta z=50$~$\mu$m, show that the acoustic field generated by our focused transducer is axisymmetric around the direction of propagation and that the acoustic power is localized in a small focal spot of diameter $0.6 \, \text{mm}$ and focal depth $2 \, \text{mm}$ at $-3 \, \text{dB}$, located at $25 \, \text{mm}$ from the vibrating surface. As reported in the lower panels of Fig.~\ref{fig:setup}(b) experimental measurements are well fitted by the following theoretical expressions for a hemispherical transducer \cite{Kino:1987} without any free parameter:
\begin{eqnarray}
\label{eq:Px}P(x,0,0)&=& P_{\rm max}\, \frac{ \lambda \ell_{\rm f}}{\pi a x}\,\mathrm{J}_1\left(\frac{2\pi a x}{\lambda \ell_{\rm f}} \right)   \,,\\
\label{eq:Pz}P(0,0,z)&=& P_{\rm max}\, \frac{\ell_{\rm f}}{z+\ell_{\rm f}}\, \mathrm{sinc}\left( \frac{a^2}{2\lambda \ell_{\rm f}} \frac{z}{z+\ell_{\rm f}} \right) \, ,
\end{eqnarray}
\noindent where  $\mathrm{J}_1$ is the Bessel function of the first kind, $a = 12.5$~mm is the radius of the transducer, $\ell_{\rm f}=25$~mm is its focal length and $\lambda=c_{\rm f}/f=300~\mu$m is the acoustic wavelength (with $c_{\rm f}=1500$~m\,s$^{-1}$ the sound speed in water at $20\,^\circ \text{C}$).

Pressure measurements in the presence of grains are displayed in Fig.~\ref{fig:setup}(c). In this case, as scanning the pressure field directly within the grains would damage the hydrophone, a couple of grain layers were deposited on the membrane located in the focal plane $z=0$ and the needle hydrophone was scanned just above the grains. We checked that the pulse duration was short enough to ensure the absence of any grain motion due to acoustic radiation forces. The pressure field of Fig.~\ref{fig:setup}(c) shows strong heterogeneity and speckle-like features: due to the acoustic impedance mismatch between the glass beads and the surrounding water, the ultrasonic incident beam gets scattered and the acoustic field behind the grain layers is strongly distorted. Although a marked increase can still be observed around the focal spot, the lower panel in Fig.~\ref{fig:setup}(c) shows that the pressure strongly drops due to the grains: it only reaches about $0.1 P_{\rm max}$ while the pressure level in the absence of grains for $z_0=1$~mm is about $0.7 P_{\rm max}$ [see Fig.~\ref{fig:setup}(b)].

We conclude from Fig.~\ref{fig:setup}(c) that when the cell is filled with grains the pressure field rapidly decays after a few grain layers due to strong scattering by the glass beads. For this practical reason we chose to locate the ultrasound focal spot within the first grain layers (at about 0.5~mm behind the membrane). In order to optimize the visualization of the grain displacements, the ultrasonic beam is focused at about 1~mm behind the wall of the cell facing the CCD camera [see Fig.~\ref{fig:setup}(a)]. In the following the intensity of the ultrasonic pulses is simply given as the acoustic pressure $P$ measured at the focal spot in the absence of grains.

\subsection{Orders of magnitude for acoustic radiation pressure and acoustic streaming} 

It is well known that HIFU can generate {\it steady} stresses in a particulate suspension via two nonlinear phenomena: (i) acoustic radiation pressure due to the acoustic impedance contrast between the particles and the surrounding fluid \cite{Yosioka:1955,Chen:1996} and (ii) acoustic streaming, i.e. the flow of the suspending fluid itself due to the absorption of the acoustic wave during its propagation \cite{Lighthill:1978,Nyborg:1998}.  

Acoustic radiation pressure results from a momentum transfer caused by reflexion and refraction of an acoustic wave on an obstacle. In the case of a plane obstacle of density $\rho_\text{o}$, sound speed $c_\text{o}$ and acoustic impedance $Z_\text{o}=\rho_\text{o} c_\text{o}$, submitted to a constant, uniform acoustic beam of pressure amplitude $P$ and denoting by $c_\text{f}$ the sound speed in the surrounding fluid, $\rho_\text{f}$ its density and $Z_\text{f}=\rho_\text{f}c_\text{f}$ its acoustic impedance, the acoustic radiation pressure reads~\cite{Chu:1982}:
\begin{equation}	
\label{eq:radiation_pressure}
\Pi_\text{rad} = \left[ 1 + \left( \frac{Z_\text{f}-Z_\text{o}}{Z_\text{f}+Z_\text{o}} \right)^2 - \frac{c_\text{f}}{c_\text{o}} \frac{4Z_\text{f} Z_\text{o}}{(Z_\text{f}+Z_\text{o})^2} \right] \,\frac{P^2}{2\rho_\text{f} c_\text{f}^2}\,.
\end{equation}
Although it only applies for a plane obstacle in a uniform acoustic beam, we can use Eq.~(\ref{eq:radiation_pressure}) to estimate the acoustic radiation force on a sphere of radius $r=d/2$ along the acoustic propagation axis as $F_\text{rad}\sim\Pi_\text{rad}\pi r^2$. With $\rho_\text{o}=2.5\,10^3$~kg\,m$^{-3}$, $c_\text{o}=5,600$~m\,s$^{-1}$ for glass and $\rho_\text{f}=10^3$~kg\,m$^{-3}$ and $c_\text{f}=1,500$~m\,s$^{-1}$ for water and taking $d=0.6$~mm and $P=1$~MPa, we find $f_\text{rad}\simeq 10^{-4}$~N i.e. $f_\text{rad}\simeq 35\,mg$ with $mg$ the average grain weight. This estimate is however obviously very crude as, in our case, (i)~the sphere has a diameter comparable to the acoustic wavelength and (ii)~the acoustic field is strongly focused over a few acoustic wavelengths. Therefore one has to take into account the exact forms of both the acoustic field radiated by the sphere and the focused incident beam. This more general case of a small elastic sphere in a focused beam has been worked out in \cite{Chen:1996}. Beside the longitudinal wave velocity, it involves the velocity $c_s$ of shear waves within the obstacle ($c_s=3,800$~m\,s$^{-1}$ for glass). Implementing the full computation in our experimental conditions leads to $f_\text{rad}\simeq 3\,mg$ for $P=1$~MPa.

Nonlinear acoustic propagation also induces attenuation which results in momentum transfer from the acoustic wave to the fluid and hence in a flow field with velocity $\vec{v}_\text{s}$ \cite{Lighthill:1978}. Modeling the acoustic beam around the spherical obstacle as a cylinder of radius $\ell_x$ and length $\ell_z$, the magnitude of $\vec{v}_\text{s}$ at focus can be estimated as~\cite{Kamakura:1995,Nyborg:1998}:
\begin{equation}
\label{eq:vs}
v_\text{s} = \frac{\omega^2\ell_x^2}{2\rho_\text{f}^2c_\text{f}^5} \left(\frac{1}{3} + \frac{\zeta}{4\eta}\right) \ln{\left(\frac{\sqrt{4\ell_x^2+\ell_z^2}+\ell_z}{\sqrt{4\ell_x^2+\ell_z^2}-\ell_z}\right)} P^2\,,
\end{equation}
where $\omega=2\pi f$ is the wave pulsation and $\eta$ and $\zeta$ are the dynamic and bulk viscosities of the fluid respectively. This phenomenon is referred to as ``acoustic streaming.'' For our experiment at $f=5$~MHz in water, with $\ell_x=0.3$~mm and $\ell_z=2$~mm, we find a typical streaming velocity of $v_\text{s}\simeq 2$~cm\,s$^{-1}$ for $P=1$~MPa. Note that this estimate does not account for the presence of bounding walls and obstacles. In practice this velocity most probably largely overestimates the actual streaming velocity which is strongly limited by both the presence of the membrane and the large compacity of the granular packing. In any case, if present, such streaming flow is the cause of a steady inertial force acting on the obstacle that reads~\cite{Maxey:1983}:
\begin{equation}
\label{eq:streaming}
f_\text{inertia}=\frac{3m_\text{f}}{2} \| (\vec{v}_\text{s} \cdot \overrightarrow{\nabla}) \vec{v}_\text{s} \|\simeq \frac{3m_\text{f} v_\text{s}^2}{2\ell_z} \,,
\end{equation}
where $m_\text{f}=\pi\rho_\text{f} d^3/6$ is the mass of the fluid displaced by the elastic sphere. Finally, for small Reynolds numbers, the streaming flow also exerts a viscous drag on the obstacle:
\begin{equation}
\label{eq:drag}
f_\text{drag} = 6\pi \eta r v_\text{s}\,.
\end{equation}

Eqs.~(\ref{eq:streaming}) and (\ref{eq:drag}) respectively yield $f_\text{inertia} \sim 0.01\,mg$ and $f_\text{drag} \sim 0.04\,mg$ for $P=1$~MPa. Therefore the forces due to acoustic streaming are about two orders of magnitude smaller than $f_\text{rad}\simeq 3\,mg$. Since $f_\text{rad}\sim P^2\sim f_\text{drag}$ and $f_\text{inertia}\sim P^4$ this remains valid over all our experimental pressure range. We conclude that acoustic radiation pressure always dominates forces due to acoustic streaming. This result is also supported by movies of the granular packing that show that the typical velocity and acceleration of the grains within the focal spot are $\sim 0.1$~m\,s$^{-1}$ and $\sim 10$~m\,s$^{-2}$ respectively, well above the (over)estimated $v_\text{s}$ and similar to the acceleration due to acoustic radiation pressure. 

\subsection{Numerical methods}

\begin{figure}[!tb]
\centering	
\includegraphics[scale=0.95]{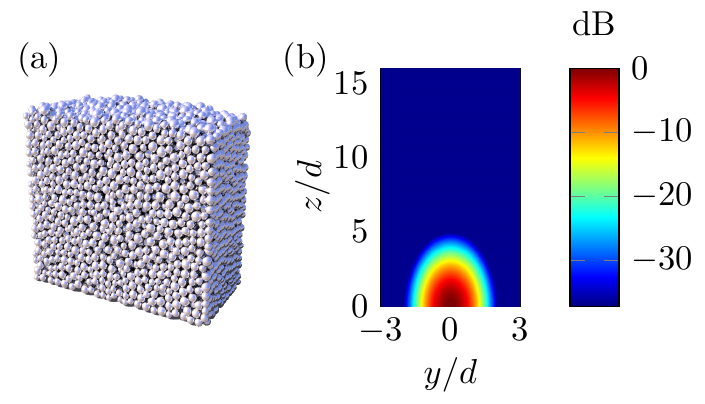}
\caption{(a)~Initial packing in DEM simulations. (b)~Force field $\mathbf{f}(x,y,z)$ used in the simulation and plotted in a logarithmic color scale (dB). $d$ is the grain average diameter.}
\label{fig:dem}
\end{figure}

In order to model the experimental situation while accessing more internal observables characteristic of the grain packing, three-dimensional numerical simulations of soft-sphere molecular dynamics, a.k.a. discrete elements method (DEM), are performed using spherical frictional grains that interact through collisions. As shown above, neglecting acoustic streaming is justified so that in the experiments interstitial water simply serves as a transmission medium for the ultrasound. Therefore the simplest numerical approach consists in simulating only dry grains. The force scheme is a linear damped spring for the normal component \cite{Schafer:1996} and a Cundall history-dependent force for the tangential component \cite{Cundall:1979}. To avoid crystallization we use polydisperse grains of density $\rho$ with average diameter $d$ and standard deviation $0.11d$ similar to experiments. Lengths, masses and accelerations are respectively normalized by $d$, $m=\pi\rho d^3/6$ and $g$ the acceleration of gravity, so that in dimensionless units the other parameters are: time step=$5.10^{-4}$, spring constant=$10^4$, friction coefficient=0.1 and restitution coefficient=0.86 (see~\cite{Percier:2013} for full technical details).

A cell of size $10d \times 20d \times 50d$ with frictional bounding walls contains up to 10,000 grains which are initially left to rest under gravity [see Fig.~\ref{fig:dem}(a)]. The acoustic radiation force is modeled by a localized force $\mathbf{f}$ applied on the center of mass of individual grains and directed upwards: $\mathbf{f}(x,y,z) = F e^{-x^2/2\sigma_x^2} e^{-y^2/2\sigma_y^2}e^{-z^2/2\sigma_z^2}\,\mathbf{e}_z$, where the origin is taken at the center of the bottom wall, $F$ is given in units of the grain weight $mg$, and $\sigma_x=\sigma_y=d$, $\sigma_z=2.5d$ which mimics the experimental focal spot  [see Fig.~\ref{fig:dem}(b)]. For comparison with experiments we define $P_{\rm DEM}\equiv\sqrt{F/mg}$ as the analogue of the pressure amplitude $P$ since acoustic radiation forces scale as $P^2$. Note that $\mathbf{f}$ is independent of the local packing fraction, which is clearly not the case in the experiment where the pressure field is coupled to the grain spatial distribution through the scattering of the incident acoustic wave as shown in Fig.~\ref{fig:setup}(c). Due to this idealized force field and to the lack of interstitial fluid, our DEM simulations are only expected to show qualitative agreement with experimental results. However they shall provide more physical insight into the effect of a localized force on the grain packing, especially on its internal structure. In order to model the experimental pulses, the driving force is switched on and off at a frequency such that the grains go back to full rest after each pulse.

\section{Results}
\label{sec:results}

\subsection{Unjamming}
\label{sec:unjamming}

\begin{figure*}[!tb]
\centering	
\includegraphics[scale=0.95]{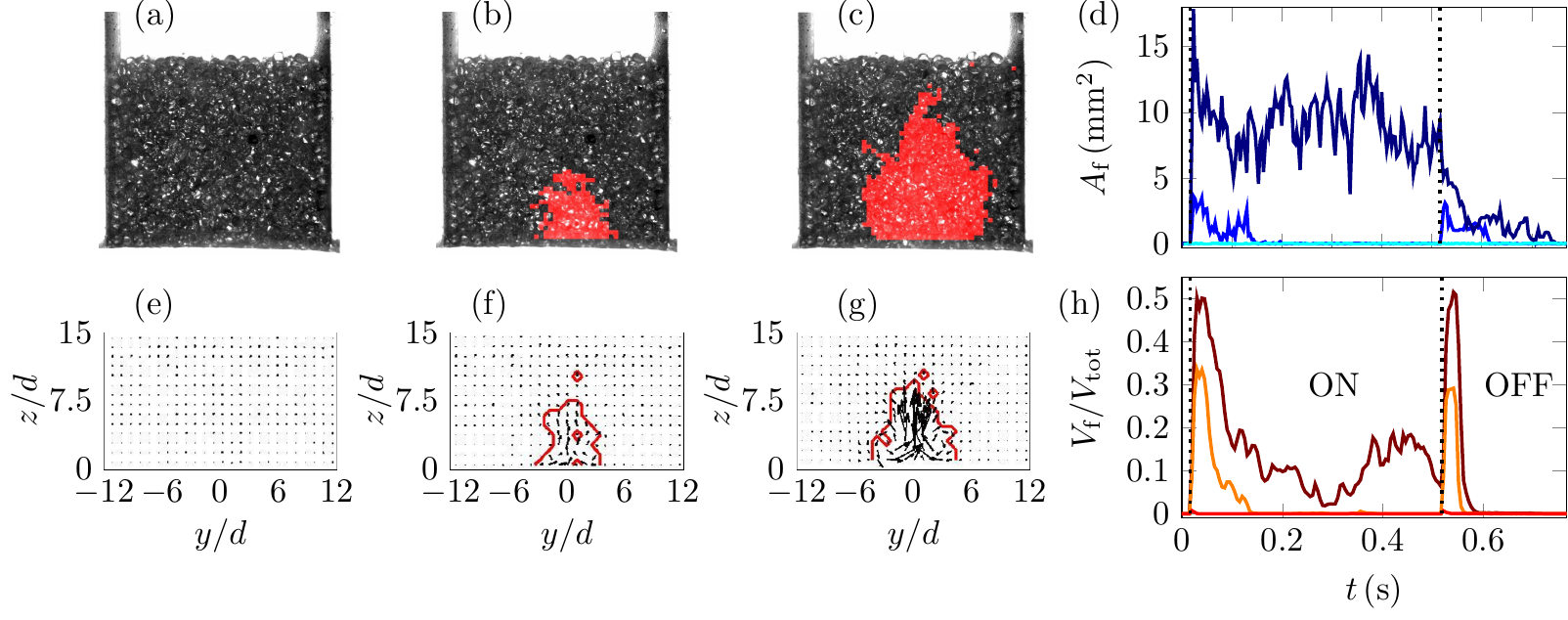}
\caption{(a,b,c)~Pictures of the granular packing highlighting (in red) the total area fluidized during one pulse of amplitude for $P=0.27$, 0.81 and 1.2~MPa respectively and (d)~fluidized area $A_{\rm f}$ versus time $t$ for the same amplitudes from bottom to top. (e,f,g)~Simulated velocity fields in the focal plane averaged over the whole pulse duration for $P_{\rm DEM}=4.4$, 9.5 and 11 respectively. The red lines surround fluidized regions. (h)~Fluidized volume $V_{\rm f}(t)$ in DEM simulations for the same amplitudes from bottom to top. $V_{\rm tot}$ is the total volume occupied by the grains and $t$ is made dimensional so as to match the experimental time. A grain is counted in $V_f$ if its displacement between two successive time steps exceeds $d/100$. See also Movies in the ESI.\dag}
\label{fig:unjamming}
\end{figure*}

In a first series of experiments, we study the effect of a single 0.5~s ultrasonic pulse on the granular packing. For low acoustic intensities the pulse has no noticeable effect [see Fig.~\ref{fig:unjamming}(a)] while for high intensities the acoustic radiation pressure unjams the pile with grains at the ultrasonic focus being pushed upwards while the surrounding grains recirculate [see Fig.~\ref{fig:unjamming}(b,c) and Movie~1 in the ESI\dag].

In order to quantify the local grain motion, we use particle image velocimetry (PIV) and compute the number of PIV cells in which the velocity exceeds the noise level (induced by image acquisition and processing) to extract the apparent ``fluidized area'' $A_{\rm f}$ as a function of time [see Fig.~\ref{fig:unjamming}(d) and Movie~2 in the ESI\dag]. At low amplitude ($P=0.27 \, \text{MPa}$), the acoustic radiation force is too small to induce rearrangements and the packing always remains in a jammed state. For an intermediate amplitude ($P=0.81 \, \text{MPa}$), close to the unjamming threshold, a few rearrangements occur at the beginning of the pulse but the pile quickly reaches a new jammed state. Finally, when $P$ is further increased ($P=1.2 \, \text{MPa}$), acoustic radiation forces induce a global yet erratic motion of the grains around the focal spot.

A similar unjamming behavior is observed in DEM simulations for the ``fluidized volume'' $V_{\rm f}$ occupied by grains with noticeable velocity [see Fig.~\ref{fig:unjamming}(e,f,g)]. As shown in Fig.~\ref{fig:unjamming}(h) however, the simulated $V_{\rm f}(t)$ signals display large peaks right after ultrasound is turned on and off. This indicates that the grains are first pushed upwards creating a small depleted zone at the focal point and then settle back after the end of the pulse (see Movies~3 and 4 in the ESI\dag). This phenomenon is much less pronounced in experiments and may be due to the absence of interstitial fluid in the simulations. In any case Fig.~\ref{fig:unjamming} clearly shows that HIFU can unjam a granular packing and set it into motion remotely and locally. This constitutes our first important result.

\subsection{Intermittency}
\label{sec:intermittency}

\begin{figure}[tb]
\centering
\includegraphics[scale=0.95]{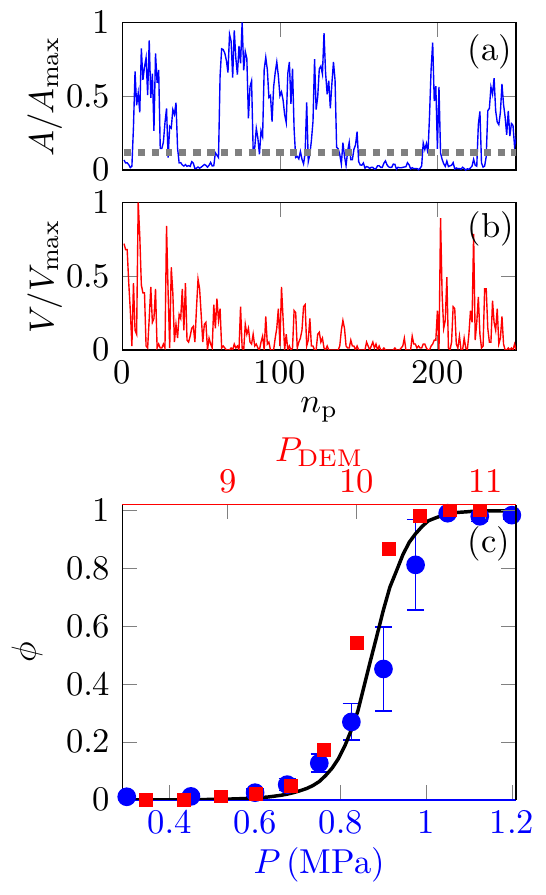}
\caption{(a)~Fluidized area $A$ normalized by its maximum value $A_{\rm max}$ vs pulse number $n_p$ for $P=0.83$~MPa. The dotted line is the threshold used for computing $\phi$. (b)~Fluidized volume $V/V_{\rm max}$ vs $n_p$ for $P_{\rm DEM}=10$. (c)~Fraction $\phi$ of pulses inducing unjamming as a function of the pulse amplitude $P$ in experiments (lower axis, blue circles) and of $P_{\rm DEM}$ in simulations (upper axis, red squares). As no direct quantitative relationship between $P$ and $P_{\rm DEM}$ can be made, the upper axis was scaled so that $\phi$ values roughly overlap. The error bars show the dispersion over different experiments. The solid line corresponds to the model discussed in Sect.~\ref{sec:model} with $\Sigma_0 = 0.525$~MPa, $\omega=0.15$~MPa, $\beta_{\rm f} = 0.02$ and $\beta_{\rm j}=0.2$.}
\label{fig:intermittency}
\end{figure}

In order to better probe the dynamics close to the unjamming threshold, we now submit the packing to long series of up to $N=1,500$ pulses of constant amplitude $P$ and duration 0.5~s, separated by rest periods of 1.5~s, and we measure the total area $A$ fluidized during each pulse. For low values of $P$ the pile is never fluidized whereas at higher $P$ every pulse induces motions in the pile. Remarkably, in the intermediate regime, the packing unjams intermittently from one pulse to another as illustrated in Fig.~\ref{fig:intermittency}(a). The simulated dynamics also show that for a given intermediate value of the amplitude, some pulses induce large-scale recirculations whereas others only trigger limited motion of individual grains which settle back to the same position after the pulse [see Fig.~\ref{fig:intermittency}(b)]. 

From such data we extract the fraction $\phi$ of pulses that induce unjamming. To discriminate between unjammed and jammed states, a noise level (due to image acquisition and processing rather than to physical causes) was extracted from images recorded without ultrasound. A pulse is counted as inducing unjamming when $A$ exceeds twice the noise level [see dotted line in Fig.~\ref{fig:intermittency}(a)]. Similarly, in DEM simulations, a pulse is counted in $\phi$ if more than two grains move by a distance larger than their diameter between their positions at rest before and after the pulse. The resulting $\phi$ vs $P$ curves are displayed in Fig.~\ref{fig:intermittency}(c). Experiments and DEM simulations show remarkably similar trends that are both well captured by the simple model introduced below in Sect.~\ref{sec:model}.

\subsection{Hysteresis}
\label{sec:hysteresis}

We finally sweep the pulse amplitude $P$ successively upwards and downwards by steps of $N=25$ pulses where $P$ is kept constant. Here again each pulse lasts 0.5~s and the packing is left at rest for 1.5~s between two pulses. For each step at fixed $P$, the fluidized area $A$ is averaged over the $N$ pulses for which the pressure amplitude is kept constant. Several cycles are performed successively on the same granular packing starting from the poorly controlled initial state that results from the sedimentation of the grains. During the first two cycles the grains gradually compact and the fluidized area decreases upon compaction. After two cycles a reproducible decreasing branch is obtained in $A(P)$, which indicates that a well-controlled packing state is reached. This latter state was taken as the initial state for all the experiments presented above. In the simulations however, no significant compaction is observed so that the initial state is always taken as a random packing resulting from simple ``rainfall'' preparation as shown in Fig.~\ref{fig:dem}(a) (see also Movies~3 and 4 in the ESI\dag).

\begin{figure}[!tb]
\centering
\includegraphics[scale=0.95]{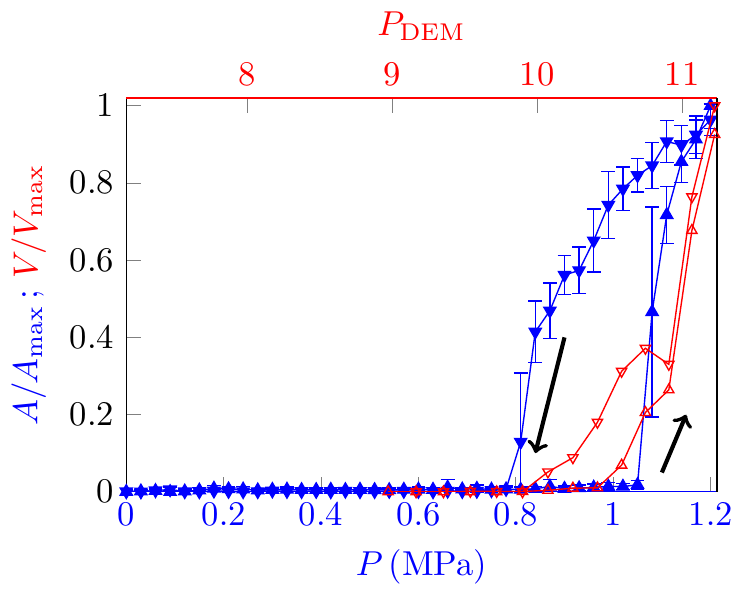}
\caption{Normalized fluidized fraction $A/A_{\rm max}$ in experiments (lower axis, blue filled symbols) and $V/V_{\rm max}$ in simulations (upper axis, red open symbols) when sweeping the pulse amplitude $P$ upwards ($\blacktriangle$) then downwards ($\blacktriangledown$). The error bars show the standard deviation over the $N=25$ successive pulses for which the pressure amplitude is kept constant. Simulations use $N=5$.}
\label{fig:hysteresis}
\end{figure}

Figure~\ref{fig:hysteresis} presents an example of a subsequent cycle and shows that unjamming under HIFU is hysteretic (see blue symbols): when decreasing $P$, the amplitude $P^-\simeq 0.75$~MPa at which ultrasonic pulses stop fluidizing the pile is significantly lower than the amplitude $P^+\simeq 1.05$~MPa at which the fluidization starts upon increasing $P$. 
While jamming along the robust decreasing branch is rather progressive, we note that unjamming upon increasing $P$ is much sharper and that $P^+$ varies from one cycle to the other, with $P^+$ only slightly larger than $P^-$ in some cases. This dispersion is consistent with the fact that intermittency is observed over the very same amplitude range $P\simeq 0.8$--1~MPa in Fig.~\ref{fig:intermittency}(c). Similar results are found for different values of $N=5$--100 or of the height of the packing (1--2~cm) as well as in DEM simulations yet with smaller hysteresis (see red symbols in Fig.~\ref{fig:hysteresis}).

\section{Discussion}
\label{sec:discuss} 

\subsection{Structural analysis of the simulated packings} 
\label{sec:chains} 

\begin{figure}[!tb]
\centering
\includegraphics[scale=0.95]{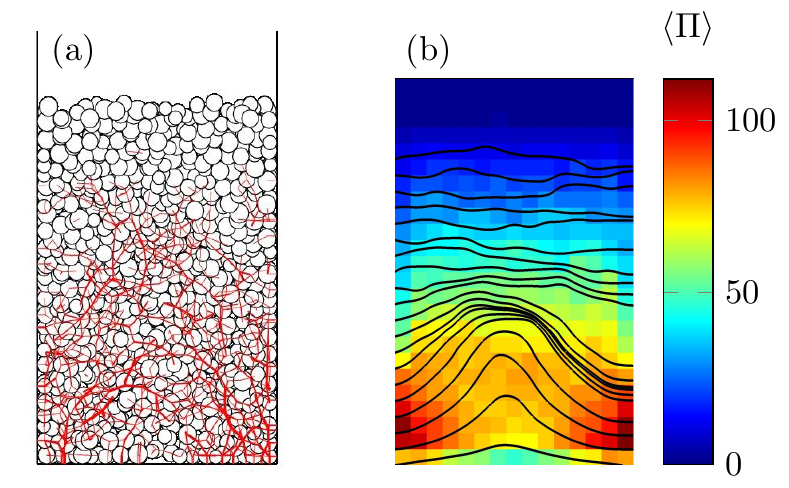}
\caption{(a)~Typical state of the simulated packing at rest between two pulses in the unjamming regime ($P_{\rm DEM}=11$) and across the focal plane $x=0$. Red lines show the force chains network with a thickness proportional to the local normal force. (b)~Map of the average pressure $\langle\Pi(\mathbf{r},n_{\rm p})\rangle$, defined as the trace of the local stress tensor and coded in linear color levels, for $P_{\rm DEM}=11$. Black lines are average streamlines of the principal direction of the stress tensor. Averages are taken over $x$ and over 200 packings at rest.\label{fig:chains}}
\end{figure}

As a first step to better grasp the physics behind the effects described above, we further analyze our DEM simulations in terms of structural parameters derived from the state of the granular packing when it is at rest between two successive pulses. Not unexpectedly, after a pulse in the unjamming regime, the packing displays large-scale arch-like structures in both the force chains and the principal stress directions (see Fig.~\ref{fig:chains}). Indeed once an acoustic pulse that fluidizes the packing is stopped, grains come back to rest but keep the fingerprint of the recirculating area around the focal spot. These arch-like structures also appear as strong spatial variations in the pressure $\Pi(\mathbf{r},n_{\rm p})$ defined as the trace of the stress tensor at point $\mathbf{r}$ once the packing has reached equilibrium after $n_{\rm p}$ pulses [see Fig.~\ref{fig:chains}(b)]. 

\begin{figure*}[!tb]
\centering
\includegraphics[scale=0.95]{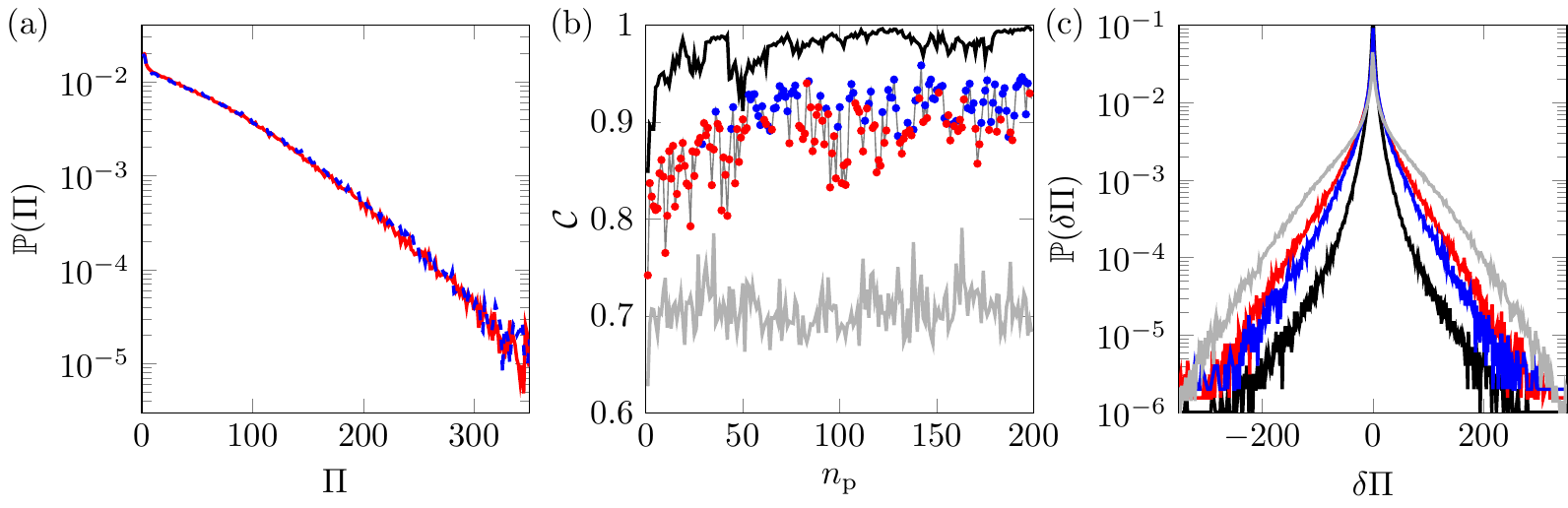}
\caption{Statistical analysis of simulated packings at rest. (a)~Probability density function of $\Pi(\mathbf{r},n_{\rm p})$ in the intermittent regime ($P_{\rm DEM}=10$) when discriminating between pulses that induce unjamming (red solid line) and pulses for which the packing remains jammed (blue dotted line). (b)~Correlation coefficient $\mathcal{C}$ between two successive pressure maps [see Eq.~(\ref{eq.correl})] and (c)~probability density function of the increments $\delta\Pi(\mathbf{r},n_{\rm p})=\Pi(\mathbf{r},n_{{\rm p}+1})-\Pi(\mathbf{r},n_{\rm p})$ for $P_{\rm DEM}=9$ (black), 10 (red and blue) and 11 (gray), respectively in the jammed, intermittent and unjamming regimes. As in (a), red corresponds to pulses inducing unjamming while blue indicates pulses for which the packing remains jammed.\label{fig:stats}}
\end{figure*}

Much more surprisingly Fig.~\ref{fig:stats}(a) shows that in the intermittent regime the distribution of $\Pi(\mathbf{r},n_{\rm p})$ taken after fluidization events is undistinguishable from that taken over packings that remain jammed. Such a similarity between the packing states is confirmed by computing the correlation coefficient $\mathcal{C}$ between two successive pressure maps at rest, defined as 
\begin{equation}
\mathcal{C}(n_{\rm p})=\frac{\langle\tilde{\Pi}(\mathbf{r},n_{\rm p})\tilde{\Pi}(\mathbf{r},n_{\rm p}+1)\rangle}{\sqrt{\langle\tilde{\Pi}^2(\mathbf{r},n_{\rm p})\rangle}\sqrt{\langle\tilde{\Pi}^2(\mathbf{r},n_{\rm p}+1)\rangle}}\,,
\label{eq.correl}
\end{equation}
where $\tilde{\Pi}(\mathbf{r},n_{\rm p})=\Pi(\mathbf{r},n_{\rm p})-\langle\Pi(\mathbf{r},n_{\rm p})\rangle$ and the average is taken over all points $\mathbf{r}$. One expects $\mathcal{C}\simeq 1$ for two identical successive packings and the more successive packings differ, the smaller the values of $\mathcal{C}$. In the intermittent regime, $\mathcal{C}$ takes values intermediate between the jammed and fluidized cases [see Fig.~\ref{fig:stats}(b)], while one would expect it to jump from one case to another if the two packing states at rest were structurally very different.

Similar results are obtained when only a small region around the focal spot is considered or when using other standard structural observables including the local number of contacts, the principal direction of the stress tensor, its anisotropy and force histograms. This implies that, at least for the simulated dry grains, it is very difficult to predict whether the system will unjam or not based on static features alone. Yet, looking more closely at $\mathcal{C}(n_{\rm p})$, one notices that the correlation coefficient is slightly smaller for pulses that induce unjamming [red points in Fig.~\ref{fig:stats}(b), $\langle C\rangle= 0.873\pm0.017$] than for pulses for which the packing remains jammed [blue points in Fig.~\ref{fig:stats}(b), $\langle C\rangle= 0.921\pm0.035$]. This subtle difference between the two groups of pulses can also be evidenced by investigating the statistics of the pressure increments from one pulse to the next one, $\delta\Pi(\mathbf{r},n_{\rm p})=\Pi(\mathbf{r},n_{{\rm p}+1})-\Pi(\mathbf{r},n_{\rm p})$. As shown in Fig.~\ref{fig:stats}(c), the distribution of $\delta\Pi(\mathbf{r},n_{\rm p})$ slightly depends on whether pulse number $n_{\rm p}$ has induced unjamming or not.
When grains remain in a jammed state, the average and standard deviation of the increments are respectively $+0.196$ and 28.8, while one gets $-0.218$ and 36.0 when fluidization occurs. This means that the local pressure tends to increase when the system remains in the jammed state (i.e. the packing is consolidated by successive pulses) while fluidization leads on average to smaller pressure levels yet with larger spatial variations (i.e. increased disorder).
We also note that in the intermittent regime, the probability density function of $\delta\Pi$ displays exponential tails that lie between the fully jammed and unjammed cases which respectively display tails that are close to power-law and Gaussian distributions. These features certainly deserve more attention and should be addressed in future work as they could provide some hints on the mechanisms by which the previous history of the packing is encoded into its structure at rest.

\subsection{Simple two-state model} 
\label{sec:model} 

The above analysis of the simulation results suggests that unjamming depends on dynamical phenomena that occur when the localized force is applied and somehow affect the subsequent structure and history-dependent response in a subtle way. Based on this observation we propose a minimal heuristic model that relies on assumptions reminiscent of trap models \cite{Bouchaud:1992} and of the Soft Glassy Rheology (SGR) model \cite{Sollich:1997}. The state of the pile is described by an internal stress $\Sigma$ representing its ability to resist shear and randomly taken with a Gaussian distribution of mean $\Sigma_0$ and standard deviation $\omega$:
\begin{equation}
\mathbb{P}(\Sigma) = \frac{1}{\sqrt{\pi} \omega} \exp{\left[- \left(\frac{\Sigma - \Sigma_0}{\omega}\right)^2 \right]}.
\label{eq.gaussian}
\end{equation}

Acoustic pulses are characterized by a stress $\sigma_{\rm p}$ that can possibly be changed in time. We consider only two behaviors of the granular packing when submitted to an ultrasonic pulse: either it is fluidized or it remains jammed. The history of the packing response to successive pulses is thus given by a sequence $(m_i)_{i \in \mathbb{N}}$ with values in $\{0,1\}$ indicating the state of the packing: if the  $i^{\rm th}$ pulse (corresponding to a stress $\sigma_{\text{p},i}$) fluidizes the pile then $m_i = 1$ while $m_i=0$ if the pile remains jammed. 

At step $i$ the following procedure is applied:
\begin{itemize}
\item if $\Sigma_{i-1}<\sigma_{\text{p},i}$ the pile is fluidized: we set $m_i=1$ and a new internal stress $\Sigma_i$ is taken from the Gaussian distribution [Eq.~(\ref{eq.gaussian})].
\item otherwise the pulse alone is not strong enough to overcome the internal stress. Nonetheless it still induces mechanical fluctuations in the pile that could allow for fluidization. Following the SGR model \cite{Sollich:1997}, we consider an Arrhenius-like probability of fluidization with an effective inverse temperature $\beta$ and an energy barrier $\Sigma_{i-1}-\sigma_{\text{p},i}$. We then draw the value of $m_i \in \{0,1\}$ with probability $\{1-p,p\}$ where $p=\exp{[-\beta(\Sigma_{i-1}-\sigma_{\text{p},i})]}$. Moreover we include a memory effect: if the pile was fluidized at the former step, fluctuations are assumed to be stronger than if the pile was previously jammed.
Therefore, when $\Sigma_{i-1}\ge \sigma_{\text{p},i}$,
\begin{itemize}
\item if the pile was fluidized at the former step ($m_{i-1}=1$), we set the effective temperature $1/\beta$ to a ``high'' value $1/\beta_{\rm f}$,
\item if the pile was jammed at the former step ($m_{i-1}=0$), the effective temperature is set to a ``low'' value $1/\beta_{\rm j}$ corresponding to smaller fluctuations.
\end{itemize}
\end{itemize}
These assumptions are consistent with the fact that the response of the packing to acoustic pulses mostly depends on its dynamical state (with some memory effect).  Indeed effective temperatures characterize the packing {\it during} the pulse only and the above model does not involve any static ingredient. More refined versions of our model could incorporate some interplay between the fluidization history and the packing structure, e.g. by making the internal stress distribution of Eq.~(\ref{eq.gaussian}) history-dependent or by including an effect of the number of pulses over which the packing has remained fluidized.

Keeping a constant value of $\sigma_{\rm p}$ the present model aims at reproducing the results on intermittency while cycling the value of $\sigma_{\rm p}$ corresponds to our hysteresis protocol. The model contains four adjustable parameters, $\Sigma_0$, $\omega$, $\beta_{\rm f}$ and $\beta_{\rm j}$, some of which can be estimated from experimental data. For instance, Fig.~\ref{fig:intermittency}(c) shows that for $P_\text{max} >  1.05$~MPa $\equiv \Sigma_\text{max}$ all pulses fluidize the granular packing. Consequently values of $\Sigma$ should be under this stress. Moreover negative values of $\Sigma$ are physically excluded. This constrains the acceptable values of $\Sigma_0$ and $\omega$. Here we choose $\Sigma_0 = 0.525$~MPa and $\omega=0.15$~MPa, which ensures a reasonably large distribution accounting for the range of internal stresses, with a low probability of drawing negative stresses or values above $\Sigma_\text{max}$ given by $(1-\mathrm{erf}[(\Sigma_\text{max} - \Sigma_0)/\omega]) \sim 10^{-6}$. Note that using a uniform distribution or a lognormal distribution for $\Sigma$ would avoid the problem of negative stress values. We checked that the predictions of the model are not significantly affected by the probability density function provided the mean and variance of $\mathbb{P}(\Sigma)$ are kept constant.

\begin{figure}[tb]
\centering
\includegraphics[scale=0.95]{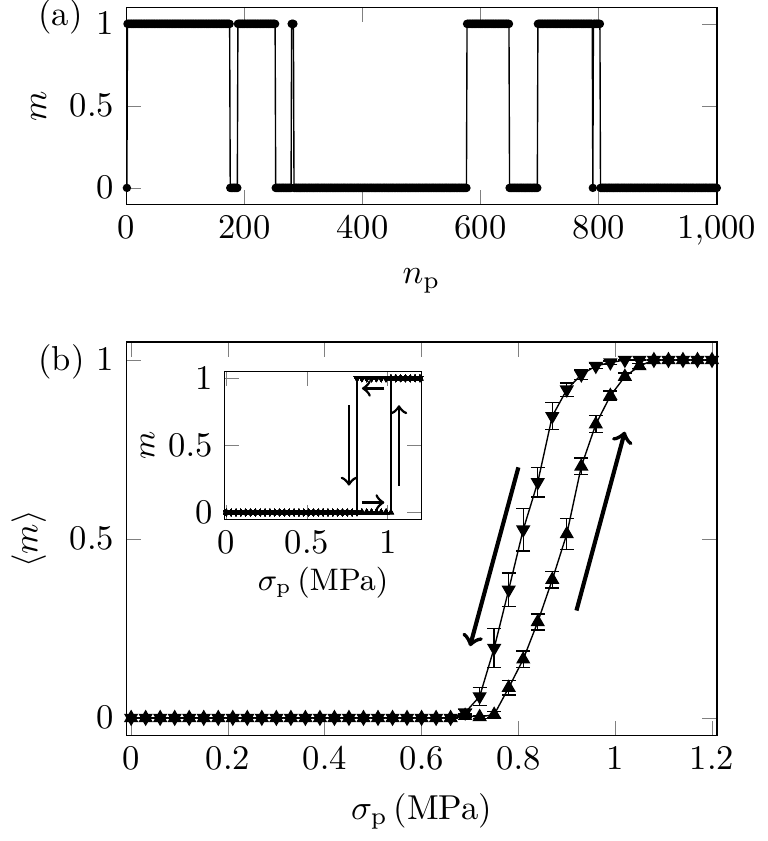}
\caption{(a)~Typical example of intermittency obtained in the model over 1,000 successive pulses. The model parameters are $\sigma_{\rm p}=0.8$~MPa, $\Sigma_0 = 0.525$~MPa, $\omega=0.15$~MPa, $\beta_{\rm f} = 0.02$ and $\beta_{\rm j}=0.2$.
(b)~Hysteresis cycle averaged over 100 realizations of the model. Inset: Typical example of hysteresis cycle obtained in a single realization of the model with the cycling protocol. Same parameters as in (a).}
\label{fig:model}
\end{figure}

The choice of $\beta_{\rm f}$ and $\beta_{\rm j}$ is a priori free but their values strongly impact the predictions of the model. Here we take $\beta_{\rm f} = 0.02$ and $\beta_{\rm j}=0.2$. With this choice of parameters, the model yields intermittency for the stationary protocol as seen in Fig.~\ref{fig:model}(a). The fraction $\phi$ of pulses inducing fluidization is shown as a solid line in Fig.~\ref{fig:intermittency}(c). In the case of the cycling protocol, hysteresis is predicted with properties similar to the experiments. Upon increasing the pulse amplitude $\sigma_{\rm p}$ the system always unjams above a threshold that is significantly larger than that obtained for jamming upon decreasing $\sigma_{\rm p}$. On the decreasing branch there is generally no new fluidization event once a jammed state has been reached. However the decreasing branch displays important variability and some intermittency is observed on the increasing branch. A typical example of a single hysteresis cycle is displayed in the inset of Fig.~\ref{fig:model}(b) for pulse amplitudes up to $1.2$~MPa, steps of $0.03$~MPa and $N=25$ pulses per step where $\sigma_{\rm p}$ is kept constant. The hysteresis cycle obtained by averaging 100 different realizations of the model is displayed in the main panel of Fig.~\ref{fig:model}(b). This last figure shows that despite important variability in the cycles there exists a clear hysteresis and typical values for jamming and unjamming thresholds are consistent with the boundaries of the intermittent regime obtained in the experiments. We note however that the number of pulses over which the system remains (un)jammed is much larger in single realizations of our model (typically hundreds of pulses) than in experiments and simulations (typically tens of pulses). More intermittent signals can be obtained by tuning the parameters of the model but the corresponding cycling protocols no longer provide clear hysteresis branches and rather remain strongly intermittent throughout the intermediate regime.

When $\beta_f$ or $\beta_j$ is increased while keeping $\mathbb{P}(\Sigma)$ and  $N$ constant, i.e. when lowering the effective temperatures, the $\phi(P)$ curves and the hysteresis cycles are observed to shift towards larger $\sigma_{\rm p}$ values but their shapes are roughly conserved. On the other hand, when $\beta_f$ or $\beta_j$ is decreased, the curves shift to smaller $\sigma_{\rm p}$ and the hysteresis eventually disappears. This is expected as larger (smaller resp.) stresses are required to unjam the system for lower (higher resp.) effective temperatures. When the number of pulses per step $N$ is increased in the cycling protocol with $\beta_{\rm f} = 0.02$ and $\beta_{\rm j}=0.2$, the hysteresis cycle keeps roughly the same amplitude but shifts to larger $\sigma_{\rm p}$, most likely because the probability of getting trapped in a deep energy minimum largely exceeds that of being fluidized again.

Finally the hysteresis cycle inferred from the model is much smoother and more symmetric than the experimental and numerical ones [compare Fig.~\ref{fig:model}(b) and Fig.~\ref{fig:hysteresis}]. This is due to the fact that Fig.~\ref{fig:model}(b) results from an average over independent realizations that are intrinsically symmetric, jumping from 0 to 1, as the material can only be either fully jammed or fully fluidized in the model. In the experiments and simulations however, only part of the system may be jammed or fluidized and local heterogeneities can be highly correlated through avalanches or long-range force chains. Capturing the asymmetry of the hysteresis certainly requires to include such local effects in the model.

\subsection{Comparison with previous studies}

Classical methods for perturbing grain assemblies include shear \cite{Pouliquen:2003,Nguyen:2011,Boyer:2011,Nichol:2010}, vertical vibrations \cite{Dijksman:2011,Hanotin:2012,Zaitsev:2014}, thermal cycling \cite{Percier:2013} and injection/suction of an interstitial fluid \cite{Schroter:2005,Johnsen:2008,Brown:2010}. They all involve a {\it global} driving of the material at the scale of the container. So far unjamming a granular packing {\it locally} had only been achieved through intrusive methods, e.g. by moving an intruder or by blowing a fluid through a granular material  \cite{Peng:1995,Kolb:2004,Zuriguel:2005,Candelier:2009,Varas:2011,MacMinn:2015}. Here we have shown that HIFU can remotely induce local fluidization through acoustic radiation pressure, in contrast with previous ultrasonic studies on granular packings that reported much weaker effects \cite{Jia:2011,Espindola:2012}.

Moreover this fluidization mode displays intermittency and hysteresis, two distinctive features that are commonly encountered in the physics of granular materials. For instance, experiments in a rotating drum have shown both features in avalanching behavior along the grain surface \cite{Rajchenbach:1990,Fischer:2009,Yang:2015}. Hysteresis is also typical of the compaction kinetics of a granular column under vertical tapping \cite{Nowak:1998,Philippe:2001} and of fluidization in tapered granular beds \cite{Peng:1995}. In the latter case hysteresis was attributed to the formation of a fluidized cavity at the bottom of the bed which compresses the jammed region when the inlet flow velocity is progressively ramped up. In the present acoustically-induced fluidization however, energy is injected discontinuously into the granular packing and the system is given enough time to come to full rest and to relax possible density heterogeneities between two pulses, so that a similar mechanism is unlikely to explain hysteresis. To build a proper comparison with previous works on fluidized beds, future HIFU experiments should focus on a continuous insonification of granular packings by using transducers that can sustain high power levels on durations much longer than 1~s without overheating.

Finally we note that arguments similar to those expressed in the heuristic model of Sect.~\ref{sec:model} have already been used in the context of granular materials by \cite{Behringer:2008,Bi:2009}. There, activated processed were invoked to describe the logarithmic strengthening of granular materials under continuous shear.

\section{Open questions and conclusion}

The present results open a number of questions regarding (i)~the state of the HIFU-induced fluidized state compared to the jammed state, (ii)~the possibility of long-range effects similar to those induced by steady shear in split-bottom cells \cite{Nichol:2010} and (iii)~the physical interpretation of intermittency and hysteresis from microscopic approaches. Throughout this article we have highlighted subtle --yet most probably physically relevant-- differences between experimental and numerical results. For instance the fact that simulated hysteresis cycles are smaller than in the experiments [see Fig.~\ref{fig:hysteresis}] hint at weaker memory effects. This would be consistent with the time series of Figs.~\ref{fig:intermittency}(a) and (b) where the experimental fluidized area $A(t)$ appears to remain correlated over a larger number of successive pulses than the simulated $V(t)$. Together with the observation of different short-time transient responses in Figs.~\ref{fig:unjamming}(d) and (h), these discrepancies between experiments and simulations point to the possible influence of the interactions between grains, of their surface roughness and/or of the interstitial fluid on the fluidization characteristics and more specifically on its history-dependence. Since friction has been recognized to crucially influence the jamming of granular packings, whether static \cite{Radjai:1996,Silbert:2002}, sheared \cite{Bi:2011,Otsuki:2011} or compressed \cite{Somfai:2007,Bandi:2013}, it can also be naturally expected to play a significant role in HIFU-induced unjamming. Therefore addressing the above issues calls for more work where the grain properties shall be systematically varied.

To conclude HIFU appears as a useful, nonintrusive tool to locally unjam granular packings. More generally it could be further used to probe the physics of jamming in assemblies of soft particles. By tuning the acoustic properties of both the fluid and the grains so that radiation pressure still dominates over streaming but with much less scattering, e.g. by using packings of hydrogel particles in glycerol--water mixtures, we can expect to target virtually any position within the bulk. This should allow us to map the local susceptibility of the packing in three dimensions and hence identify stronger/weaker zones within the jammed material. Such heterogeneities could then be analyzed in terms of local correlation lengths as a function of the initial packing fraction, in an effort to address the long-standing issue of diverging scales upon three-dimensional jamming \cite{Silbert:2005,Mailman:2012}. Finally, on the applicative side, HIFU-induced unjamming could prove relevant for practical situations where fluidizing granular sediments remotely and locally is required, e.g. for unclogging or filtration problems.

\section*{Acknowledgments}   
The authors thank T.~Divoux, Y.~Forterre, B.~Issenmann, O.~Pouliquen and R.~Wunenburger for fruitful discussions. This work was funded by the Institut Universitaire de France and by the European Research Council under the European Union's Seventh Framework Programme (FP7/2007-2013) / ERC grant agreement No.~258803. 

\section*{Supplementary movies\dag}

Movie~1 presents the evolution of our granular packing during a single high-intensity focused ultrasonic pulse of duration 0.5~s and pressure amplitude $P=1.2$~MPa. Images are recorded at 300 fps. The granular packing is first submitted to two cycles of pulses with increasing then decreasing amplitude in order to start from a reproducible initial state.

Movie~2 shows the results of image analysis from Movie~1 based on particle image velocimetry. Grains with an instantaneous velocity larger than $1.2$~mm.s$^{-1}$ are counted as ``fluidized'' and the corresponding PIV cells are colored in red. This allows one to directly visualize the fluidized area $A_{\rm f}$ on the original images. The resulting $A_{\rm f}(t)$ signal is shown as the upper curve in Fig.~\ref{fig:unjamming}(d).

Movie~3 displays the results of our DEM numerical simulation for $P_{\rm DEM}=11$. The packing is first prepared by letting about 7,500 grains settle under gravity. Then a cut through the middle of the cell is shown to emphasize the localized effect of ultrasound on the grains in the bulk. This simulation corresponds to the topmost curve in Fig.~\ref{fig:unjamming}(h).

Movie~4 shows the velocity field inferred from the DEM simulation shown in Movie~3. Velocities are computed by tracking the motion of the grains located within the middle plane of the simulation box i.e. at $x=0$.

\footnotesize{

\begin{mcitethebibliography}{67}
\providecommand*{\natexlab}[1]{#1}
\providecommand*{\mciteSetBstSublistMode}[1]{}
\providecommand*{\mciteSetBstMaxWidthForm}[2]{}
\providecommand*{\mciteBstWouldAddEndPuncttrue}
  {\def\EndOfBibitem{\unskip.}}
\providecommand*{\mciteBstWouldAddEndPunctfalse}
  {\let\EndOfBibitem\relax}
\providecommand*{\mciteSetBstMidEndSepPunct}[3]{}
\providecommand*{\mciteSetBstSublistLabelBeginEnd}[3]{}
\providecommand*{\EndOfBibitem}{}
\mciteSetBstSublistMode{f}
\mciteSetBstMaxWidthForm{subitem}
{(\emph{\alph{mcitesubitemcount}})}
\mciteSetBstSublistLabelBeginEnd{\mcitemaxwidthsubitemform\space}
{\relax}{\relax}

\bibitem[Kennedy(2005)]{Kennedy:2005}
J.~Kennedy, \emph{Nat. Rev. Cancer}, 2005, \textbf{5}, 321--327\relax
\mciteBstWouldAddEndPuncttrue
\mciteSetBstMidEndSepPunct{\mcitedefaultmidpunct}
{\mcitedefaultendpunct}{\mcitedefaultseppunct}\relax
\EndOfBibitem
\bibitem[ter Haar(2007)]{terHaar:2007}
G.~ter Haar, \emph{Prog. Biophys. Mol. Biol.}, 2007, \textbf{93},
  111--129\relax
\mciteBstWouldAddEndPuncttrue
\mciteSetBstMidEndSepPunct{\mcitedefaultmidpunct}
{\mcitedefaultendpunct}{\mcitedefaultseppunct}\relax
\EndOfBibitem
\bibitem[Bercoff \emph{et~al.}(2004)Bercoff, Tanter, and Fink]{Bercoff:2004b}
J.~Bercoff, M.~Tanter and M.~Fink, \emph{IEEE Trans. Ultrason. Ferroelec. Freq.
  Contr.}, 2004, \textbf{51}, 396--409\relax
\mciteBstWouldAddEndPuncttrue
\mciteSetBstMidEndSepPunct{\mcitedefaultmidpunct}
{\mcitedefaultendpunct}{\mcitedefaultseppunct}\relax
\EndOfBibitem
\bibitem[Marmottant and Hilgenfeldt(2003)]{Marmottant:2003}
P.~Marmottant and S.~Hilgenfeldt, \emph{Nature}, 2003, \textbf{423},
  153--156\relax
\mciteBstWouldAddEndPuncttrue
\mciteSetBstMidEndSepPunct{\mcitedefaultmidpunct}
{\mcitedefaultendpunct}{\mcitedefaultseppunct}\relax
\EndOfBibitem
\bibitem[Prentice \emph{et~al.}(2005)Prentice, Cuschieri, Dholakia, Prausnitz,
  and Campbell]{Prentice:2005}
P.~Prentice, A.~Cuschieri, K.~Dholakia, M.~Prausnitz and P.~Campbell,
  \emph{Nature Phys.}, 2005, \textbf{1}, 107--110\relax
\mciteBstWouldAddEndPuncttrue
\mciteSetBstMidEndSepPunct{\mcitedefaultmidpunct}
{\mcitedefaultendpunct}{\mcitedefaultseppunct}\relax
\EndOfBibitem
\bibitem[Cates \emph{et~al.}(1998)Cates, Wittmer, Bouchaud, and
  Claudin]{Cates:1998}
M.~E. Cates, J.~P. Wittmer, J.-P. Bouchaud and P.~Claudin, \emph{Phys. Rev.
  Lett.}, 1998, \textbf{81}, 1841--1844\relax
\mciteBstWouldAddEndPuncttrue
\mciteSetBstMidEndSepPunct{\mcitedefaultmidpunct}
{\mcitedefaultendpunct}{\mcitedefaultseppunct}\relax
\EndOfBibitem
\bibitem[Liu and Nagel(1998)]{Liu:1998}
A.~Liu and S.~R. Nagel, \emph{Nature}, 1998, \textbf{396}, 21--22\relax
\mciteBstWouldAddEndPuncttrue
\mciteSetBstMidEndSepPunct{\mcitedefaultmidpunct}
{\mcitedefaultendpunct}{\mcitedefaultseppunct}\relax
\EndOfBibitem
\bibitem[Olsson and Teitel(2007)]{Olsson:2007}
P.~Olsson and S.~Teitel, \emph{Phys. Rev. Lett.}, 2007, \textbf{99},
  178001\relax
\mciteBstWouldAddEndPuncttrue
\mciteSetBstMidEndSepPunct{\mcitedefaultmidpunct}
{\mcitedefaultendpunct}{\mcitedefaultseppunct}\relax
\EndOfBibitem
\bibitem[Liu and Nagel(2010)]{Liu:2010}
A.~Liu and S.~R. Nagel, \emph{Annu. Rev. Cond. Matter Phys.}, 2010, \textbf{1},
  347--369\relax
\mciteBstWouldAddEndPuncttrue
\mciteSetBstMidEndSepPunct{\mcitedefaultmidpunct}
{\mcitedefaultendpunct}{\mcitedefaultseppunct}\relax
\EndOfBibitem
\bibitem[Ikeda \emph{et~al.}(2012)Ikeda, Berthier, and Sollich]{Ikeda:2012}
A.~Ikeda, L.~Berthier and P.~Sollich, \emph{Phys. Rev. Lett.}, 2012,
  \textbf{109}, 018301\relax
\mciteBstWouldAddEndPuncttrue
\mciteSetBstMidEndSepPunct{\mcitedefaultmidpunct}
{\mcitedefaultendpunct}{\mcitedefaultseppunct}\relax
\EndOfBibitem
\bibitem[Lerner \emph{et~al.}(2012)Lerner, During, and Wyart]{Lerner:2012a}
E.~Lerner, G.~During and M.~Wyart, \emph{Proc. Natl. Acad. Sci. USA}, 2012,
  \textbf{109}, 4798--4803\relax
\mciteBstWouldAddEndPuncttrue
\mciteSetBstMidEndSepPunct{\mcitedefaultmidpunct}
{\mcitedefaultendpunct}{\mcitedefaultseppunct}\relax
\EndOfBibitem
\bibitem[Howell \emph{et~al.}(1999)Howell, Behringer, and Veje]{Howell:1999}
D.~Howell, R.~Behringer and C.~Veje, \emph{Phys. Rev. Lett.}, 1999,
  \textbf{82}, 5241--5244\relax
\mciteBstWouldAddEndPuncttrue
\mciteSetBstMidEndSepPunct{\mcitedefaultmidpunct}
{\mcitedefaultendpunct}{\mcitedefaultseppunct}\relax
\EndOfBibitem
\bibitem[Silbert \emph{et~al.}(2005)Silbert, Liu, and Nagel]{Silbert:2005}
L.~Silbert, A.~Liu and S.~Nagel, \emph{Phys. Rev. Lett.}, 2005, \textbf{95},
  098301\relax
\mciteBstWouldAddEndPuncttrue
\mciteSetBstMidEndSepPunct{\mcitedefaultmidpunct}
{\mcitedefaultendpunct}{\mcitedefaultseppunct}\relax
\EndOfBibitem
\bibitem[Berthier \emph{et~al.}(2005)Berthier, Biroli, Bouchaud, Cipelletti,
  Masri, L'H{\^o}te, Ladieu, and Pierno]{Berthier:2005}
L.~Berthier, G.~Biroli, J.-P. Bouchaud, L.~Cipelletti, D.~E. Masri,
  D.~L'H{\^o}te, F.~Ladieu and M.~Pierno, \emph{Science}, 2005, \textbf{310},
  1797--1800\relax
\mciteBstWouldAddEndPuncttrue
\mciteSetBstMidEndSepPunct{\mcitedefaultmidpunct}
{\mcitedefaultendpunct}{\mcitedefaultseppunct}\relax
\EndOfBibitem
\bibitem[Keys \emph{et~al.}(2007)Keys, Abate, Glotzer, and Durian]{Keys:2007}
A.~S. Keys, A.~R. Abate, S.~C. Glotzer and D.~J. Durian, \emph{Nature Phys.},
  2007, \textbf{3}, 260--264\relax
\mciteBstWouldAddEndPuncttrue
\mciteSetBstMidEndSepPunct{\mcitedefaultmidpunct}
{\mcitedefaultendpunct}{\mcitedefaultseppunct}\relax
\EndOfBibitem
\bibitem[Jaeger(2015)]{Jaeger:2015}
H.~Jaeger, \emph{Soft Matter}, 2015, \textbf{11}, 12--27\relax
\mciteBstWouldAddEndPuncttrue
\mciteSetBstMidEndSepPunct{\mcitedefaultmidpunct}
{\mcitedefaultendpunct}{\mcitedefaultseppunct}\relax
\EndOfBibitem
\bibitem[Bruji{\'c} \emph{et~al.}(2005)Bruji{\'c}, Wang, Song, Johnson, Sindt,
  and Makse]{Brujic:2005}
J.~Bruji{\'c}, P.~Wang, C.~Song, D.~L. Johnson, O.~Sindt and H.~A. Makse,
  \emph{Phys. Rev. Lett.}, 2005, \textbf{95}, 128001\relax
\mciteBstWouldAddEndPuncttrue
\mciteSetBstMidEndSepPunct{\mcitedefaultmidpunct}
{\mcitedefaultendpunct}{\mcitedefaultseppunct}\relax
\EndOfBibitem
\bibitem[Dauchot \emph{et~al.}(2005)Dauchot, Marty, and Biroli]{Dauchot:2005}
O.~Dauchot, G.~Marty and G.~Biroli, \emph{Phys. Rev. Lett.}, 2005, \textbf{95},
  265701\relax
\mciteBstWouldAddEndPuncttrue
\mciteSetBstMidEndSepPunct{\mcitedefaultmidpunct}
{\mcitedefaultendpunct}{\mcitedefaultseppunct}\relax
\EndOfBibitem
\bibitem[Majmudar \emph{et~al.}(2007)Majmudar, Sperl, Luding, and
  Behringer]{Majmudar:2007}
T.~Majmudar, M.~Sperl, S.~Luding and R.~Behringer, \emph{Phys. Rev. Lett.},
  2007, \textbf{98}, 058001\relax
\mciteBstWouldAddEndPuncttrue
\mciteSetBstMidEndSepPunct{\mcitedefaultmidpunct}
{\mcitedefaultendpunct}{\mcitedefaultseppunct}\relax
\EndOfBibitem
\bibitem[Behringer \emph{et~al.}(2008)Behringer, Bi, Chakraborty, Henkes, and
  Hartley]{Behringer:2008}
R.~Behringer, D.~Bi, B.~Chakraborty, S.~Henkes and R.~Hartley, \emph{Phys. Rev.
  Lett.}, 2008, \textbf{101}, 268301\relax
\mciteBstWouldAddEndPuncttrue
\mciteSetBstMidEndSepPunct{\mcitedefaultmidpunct}
{\mcitedefaultendpunct}{\mcitedefaultseppunct}\relax
\EndOfBibitem
\bibitem[Bi \emph{et~al.}(2011)Bi, Zhang, Chakraborty, and Behringer]{Bi:2011}
D.~Bi, J.~Zhang, B.~Chakraborty and R.~Behringer, \emph{Nature}, 2011,
  \textbf{480}, 355--358\relax
\mciteBstWouldAddEndPuncttrue
\mciteSetBstMidEndSepPunct{\mcitedefaultmidpunct}
{\mcitedefaultendpunct}{\mcitedefaultseppunct}\relax
\EndOfBibitem
\bibitem[Otsuki and Hayakawa(2011)]{Otsuki:2011}
M.~Otsuki and H.~Hayakawa, \emph{Phys. Rev. E}, 2011, \textbf{83}, 051301\relax
\mciteBstWouldAddEndPuncttrue
\mciteSetBstMidEndSepPunct{\mcitedefaultmidpunct}
{\mcitedefaultendpunct}{\mcitedefaultseppunct}\relax
\EndOfBibitem
\bibitem[Somfai \emph{et~al.}(2007)Somfai, van Hecke, Ellenbroek, Shundyak, and
  van Saarloos]{Somfai:2007}
E.~Somfai, M.~van Hecke, W.~G. Ellenbroek, K.~Shundyak and W.~van Saarloos,
  \emph{Phys. Rev. E}, 2007, \textbf{75}, 020301\relax
\mciteBstWouldAddEndPuncttrue
\mciteSetBstMidEndSepPunct{\mcitedefaultmidpunct}
{\mcitedefaultendpunct}{\mcitedefaultseppunct}\relax
\EndOfBibitem
\bibitem[Bandi \emph{et~al.}(2013)Bandi, Rivera, Krzkala, and Ecke]{Bandi:2013}
M.~Bandi, M.~Rivera, F.~Krzkala and R.~Ecke, \emph{Phys. Rev. E}, 2013,
  \textbf{87}, 042205\relax
\mciteBstWouldAddEndPuncttrue
\mciteSetBstMidEndSepPunct{\mcitedefaultmidpunct}
{\mcitedefaultendpunct}{\mcitedefaultseppunct}\relax
\EndOfBibitem
\bibitem[Jia \emph{et~al.}(2011)Jia, Brunet, and Laurent]{Jia:2011}
X.~Jia, T.~Brunet and J.~Laurent, \emph{Phys. Rev. E}, 2011, \textbf{84},
  020301\relax
\mciteBstWouldAddEndPuncttrue
\mciteSetBstMidEndSepPunct{\mcitedefaultmidpunct}
{\mcitedefaultendpunct}{\mcitedefaultseppunct}\relax
\EndOfBibitem
\bibitem[Esp{\'\i}ndola \emph{et~al.}(2012)Esp{\'\i}ndola, Galaz, and
  Melo]{Espindola:2012}
D.~Esp{\'\i}ndola, B.~Galaz and F.~Melo, \emph{Phys. Rev. Lett.}, 2012,
  \textbf{109}, 158301\relax
\mciteBstWouldAddEndPuncttrue
\mciteSetBstMidEndSepPunct{\mcitedefaultmidpunct}
{\mcitedefaultendpunct}{\mcitedefaultseppunct}\relax
\EndOfBibitem
\bibitem[G{\'o}mez \emph{et~al.}(2012)G{\'o}mez, Turner, van Hecke, and
  Vitelli]{Gomez:2012}
L.~G{\'o}mez, A.~Turner, M.~van Hecke and V.~Vitelli, \emph{Phys. Rev. Lett.},
  2012, \textbf{108}, 058001\relax
\mciteBstWouldAddEndPuncttrue
\mciteSetBstMidEndSepPunct{\mcitedefaultmidpunct}
{\mcitedefaultendpunct}{\mcitedefaultseppunct}\relax
\EndOfBibitem
\bibitem[van~den Wildenberg \emph{et~al.}(2013)van~den Wildenberg, van Loo, and
  van Hecke]{vandenWildenberg:2013a}
S.~van~den Wildenberg, R.~van Loo and M.~van Hecke, \emph{Phys. Rev. Lett.},
  2013, \textbf{111}, 218003\relax
\mciteBstWouldAddEndPuncttrue
\mciteSetBstMidEndSepPunct{\mcitedefaultmidpunct}
{\mcitedefaultendpunct}{\mcitedefaultseppunct}\relax
\EndOfBibitem
\bibitem[van~den Wildenberg \emph{et~al.}(2013)van~den Wildenberg, van Hecke,
  and Jia]{vandenWildenberg:2013b}
S.~van~den Wildenberg, M.~van Hecke and X.~Jia, \emph{Europhys. Lett.}, 2013,
  \textbf{101}, 14004\relax
\mciteBstWouldAddEndPuncttrue
\mciteSetBstMidEndSepPunct{\mcitedefaultmidpunct}
{\mcitedefaultendpunct}{\mcitedefaultseppunct}\relax
\EndOfBibitem
\bibitem[Yosioka and Kawasima(1955)]{Yosioka:1955}
K.~Yosioka and Y.~Kawasima, \emph{Acustica}, 1955, \textbf{5}, 167--173\relax
\mciteBstWouldAddEndPuncttrue
\mciteSetBstMidEndSepPunct{\mcitedefaultmidpunct}
{\mcitedefaultendpunct}{\mcitedefaultseppunct}\relax
\EndOfBibitem
\bibitem[Chen and Apfel(1996)]{Chen:1996}
X.~Chen and R.~E. Apfel, \emph{J. Acoust. Soc. Am.}, 1996, \textbf{99},
  713--724\relax
\mciteBstWouldAddEndPuncttrue
\mciteSetBstMidEndSepPunct{\mcitedefaultmidpunct}
{\mcitedefaultendpunct}{\mcitedefaultseppunct}\relax
\EndOfBibitem
\bibitem[Kino(1987)]{Kino:1987}
G.~S. Kino, \emph{Acoustic Waves, Devices, Imaging and Ana\-log Signal
  Processing}, Prentice-Hall, Engelwood Cliffs, NJ, 1987\relax
\mciteBstWouldAddEndPuncttrue
\mciteSetBstMidEndSepPunct{\mcitedefaultmidpunct}
{\mcitedefaultendpunct}{\mcitedefaultseppunct}\relax
\EndOfBibitem
\bibitem[Lighthill(1978)]{Lighthill:1978}
J.~Lighthill, \emph{J. Sound Vib.}, 1978, \textbf{61}, 391--418\relax
\mciteBstWouldAddEndPuncttrue
\mciteSetBstMidEndSepPunct{\mcitedefaultmidpunct}
{\mcitedefaultendpunct}{\mcitedefaultseppunct}\relax
\EndOfBibitem
\bibitem[Nyborg(1998)]{Nyborg:1998}
W.~L. Nyborg, Nonlinear Acoustics, 1998, pp. 214--216\relax
\mciteBstWouldAddEndPuncttrue
\mciteSetBstMidEndSepPunct{\mcitedefaultmidpunct}
{\mcitedefaultendpunct}{\mcitedefaultseppunct}\relax
\EndOfBibitem
\bibitem[Chu and Apfel(1982)]{Chu:1982}
B.~Chu and R.~E. Apfel, \emph{J. Acoust. Soc. Am.}, 1982, \textbf{72},
  1673--1687\relax
\mciteBstWouldAddEndPuncttrue
\mciteSetBstMidEndSepPunct{\mcitedefaultmidpunct}
{\mcitedefaultendpunct}{\mcitedefaultseppunct}\relax
\EndOfBibitem
\bibitem[Kamakura \emph{et~al.}(1995)Kamakura, Matsuda, Kumamoto, and
  Breazeale]{Kamakura:1995}
T.~Kamakura, K.~Matsuda, Y.~Kumamoto and M.~A. Breazeale, \emph{J. Acoust. Soc.
  Am.}, 1995, \textbf{97}, 2740--2746\relax
\mciteBstWouldAddEndPuncttrue
\mciteSetBstMidEndSepPunct{\mcitedefaultmidpunct}
{\mcitedefaultendpunct}{\mcitedefaultseppunct}\relax
\EndOfBibitem
\bibitem[Maxey and Riley(1983)]{Maxey:1983}
M.~R. Maxey and J.~J. Riley, \emph{Phys. Fluids}, 1983, \textbf{26},
  883--889\relax
\mciteBstWouldAddEndPuncttrue
\mciteSetBstMidEndSepPunct{\mcitedefaultmidpunct}
{\mcitedefaultendpunct}{\mcitedefaultseppunct}\relax
\EndOfBibitem
\bibitem[Sch{\"a}fer \emph{et~al.}(1996)Sch{\"a}fer, Dippel, and
  Wolf]{Schafer:1996}
J.~Sch{\"a}fer, S.~Dippel and D.~Wolf, \emph{J. Phys. I}, 1996, \textbf{6},
  5--20\relax
\mciteBstWouldAddEndPuncttrue
\mciteSetBstMidEndSepPunct{\mcitedefaultmidpunct}
{\mcitedefaultendpunct}{\mcitedefaultseppunct}\relax
\EndOfBibitem
\bibitem[Cundall and Strack(1979)]{Cundall:1979}
P.~Cundall and O.~Strack, \emph{G{\'e}otechnique}, 1979, \textbf{29},
  47--65\relax
\mciteBstWouldAddEndPuncttrue
\mciteSetBstMidEndSepPunct{\mcitedefaultmidpunct}
{\mcitedefaultendpunct}{\mcitedefaultseppunct}\relax
\EndOfBibitem
\bibitem[Percier \emph{et~al.}(2013)Percier, Divoux, and
  Taberlet]{Percier:2013}
B.~Percier, T.~Divoux and N.~Taberlet, \emph{Europhys. Lett.}, 2013,
  \textbf{104}, 24001\relax
\mciteBstWouldAddEndPuncttrue
\mciteSetBstMidEndSepPunct{\mcitedefaultmidpunct}
{\mcitedefaultendpunct}{\mcitedefaultseppunct}\relax
\EndOfBibitem
\bibitem[Bouchaud(1992)]{Bouchaud:1992}
J.-P. Bouchaud, \emph{J. Phys. I France}, 1992, \textbf{2}, 1705--1713\relax
\mciteBstWouldAddEndPuncttrue
\mciteSetBstMidEndSepPunct{\mcitedefaultmidpunct}
{\mcitedefaultendpunct}{\mcitedefaultseppunct}\relax
\EndOfBibitem
\bibitem[Sollich \emph{et~al.}(1997)Sollich, Lequeux, H\'ebraud, and
  Cates]{Sollich:1997}
P.~Sollich, F.~Lequeux, P.~H\'ebraud and M.~E. Cates, \emph{Phys. Rev. Lett.},
  1997, \textbf{78}, 2020--2023\relax
\mciteBstWouldAddEndPuncttrue
\mciteSetBstMidEndSepPunct{\mcitedefaultmidpunct}
{\mcitedefaultendpunct}{\mcitedefaultseppunct}\relax
\EndOfBibitem
\bibitem[Pouliquen \emph{et~al.}(2003)Pouliquen, Belzons, and
  Nicolas]{Pouliquen:2003}
O.~Pouliquen, M.~Belzons and M.~Nicolas, \emph{Phys. Rev. Lett.}, 2003,
  \textbf{91}, 014301\relax
\mciteBstWouldAddEndPuncttrue
\mciteSetBstMidEndSepPunct{\mcitedefaultmidpunct}
{\mcitedefaultendpunct}{\mcitedefaultseppunct}\relax
\EndOfBibitem
\bibitem[Nguyen \emph{et~al.}(2011)Nguyen, Darnige, Bruand, and
  Clement]{Nguyen:2011}
V.~Nguyen, T.~Darnige, A.~Bruand and E.~Clement, \emph{Phys. Rev. Lett.}, 2011,
  \textbf{107}, 138303\relax
\mciteBstWouldAddEndPuncttrue
\mciteSetBstMidEndSepPunct{\mcitedefaultmidpunct}
{\mcitedefaultendpunct}{\mcitedefaultseppunct}\relax
\EndOfBibitem
\bibitem[Boyer \emph{et~al.}(2011)Boyer, Guazzelli, and Pouliquen]{Boyer:2011}
F.~Boyer, E.~Guazzelli and O.~Pouliquen, \emph{Phys. Rev. Lett.}, 2011,
  \textbf{107}, 188301\relax
\mciteBstWouldAddEndPuncttrue
\mciteSetBstMidEndSepPunct{\mcitedefaultmidpunct}
{\mcitedefaultendpunct}{\mcitedefaultseppunct}\relax
\EndOfBibitem
\bibitem[Nichol \emph{et~al.}(2010)Nichol, Zanin, Bastien, Wandersman, and van
  Hecke]{Nichol:2010}
K.~Nichol, A.~Zanin, R.~Bastien, E.~Wandersman and M.~van Hecke, \emph{Phys.
  Rev. Lett.}, 2010, \textbf{104}, 078302\relax
\mciteBstWouldAddEndPuncttrue
\mciteSetBstMidEndSepPunct{\mcitedefaultmidpunct}
{\mcitedefaultendpunct}{\mcitedefaultseppunct}\relax
\EndOfBibitem
\bibitem[Dijksman \emph{et~al.}(2011)Dijksman, Wortel, van Dellen, Dauchot, and
  van Hecke]{Dijksman:2011}
J.~Dijksman, G.~Wortel, L.~van Dellen, O.~Dauchot and M.~van Hecke, \emph{Phys.
  Rev. Lett.}, 2011, \textbf{107}, 108303\relax
\mciteBstWouldAddEndPuncttrue
\mciteSetBstMidEndSepPunct{\mcitedefaultmidpunct}
{\mcitedefaultendpunct}{\mcitedefaultseppunct}\relax
\EndOfBibitem
\bibitem[Hanotin \emph{et~al.}(2012)Hanotin, Kiesgen~de Richter, Marchal,
  Michot, and Baravian]{Hanotin:2012}
C.~Hanotin, S.~Kiesgen~de Richter, P.~Marchal, L.~Michot and C.~Baravian,
  \emph{Phys. Rev. Lett.}, 2012, \textbf{108}, 198301\relax
\mciteBstWouldAddEndPuncttrue
\mciteSetBstMidEndSepPunct{\mcitedefaultmidpunct}
{\mcitedefaultendpunct}{\mcitedefaultseppunct}\relax
\EndOfBibitem
\bibitem[Zaitsev \emph{et~al.}(2014)Zaitsev, Gusev, Tournat, and
  Richard]{Zaitsev:2014}
V.~Zaitsev, V.~Gusev, V.~Tournat and P.~Richard, \emph{Phys. Rev. Lett.}, 2014,
  \textbf{112}, 108302\relax
\mciteBstWouldAddEndPuncttrue
\mciteSetBstMidEndSepPunct{\mcitedefaultmidpunct}
{\mcitedefaultendpunct}{\mcitedefaultseppunct}\relax
\EndOfBibitem
\bibitem[Schroter \emph{et~al.}(2005)Schroter, Goldman, and
  Swinney]{Schroter:2005}
M.~Schroter, D.~Goldman and H.~Swinney, \emph{Phys. Rev. E}, 2005, \textbf{71},
  030301\relax
\mciteBstWouldAddEndPuncttrue
\mciteSetBstMidEndSepPunct{\mcitedefaultmidpunct}
{\mcitedefaultendpunct}{\mcitedefaultseppunct}\relax
\EndOfBibitem
\bibitem[Johnsen \emph{et~al.}(2008)Johnsen, Chevalier, Lindner, Toussaint,
  Cl\'ement, M\aa{}l\o{}y, Flekk\o{}y, and Schmittbuhl]{Johnsen:2008}
O.~Johnsen, C.~Chevalier, A.~Lindner, R.~Toussaint, E.~Cl\'ement,
  K.~M\aa{}l\o{}y, E.~Flekk\o{}y and J.~Schmittbuhl, \emph{Phys. Rev. E}, 2008,
  \textbf{78}, 051302\relax
\mciteBstWouldAddEndPuncttrue
\mciteSetBstMidEndSepPunct{\mcitedefaultmidpunct}
{\mcitedefaultendpunct}{\mcitedefaultseppunct}\relax
\EndOfBibitem
\bibitem[Brown \emph{et~al.}(2010)Brown, Rodenberg, Amend, Mozeika, Steltz,
  Zakin, Lipson, and Jaeger]{Brown:2010}
E.~Brown, N.~Rodenberg, J.~Amend, A.~Mozeika, E.~Steltz, M.~R. Zakin, H.~Lipson
  and H.~M. Jaeger, \emph{Proc. Natl. Acad. Sci. USA}, 2010, \textbf{107},
  18809--18814\relax
\mciteBstWouldAddEndPuncttrue
\mciteSetBstMidEndSepPunct{\mcitedefaultmidpunct}
{\mcitedefaultendpunct}{\mcitedefaultseppunct}\relax
\EndOfBibitem
\bibitem[Peng and Fan(1995)]{Peng:1995}
Y.~Peng and L.~T. Fan, \emph{Chem. Eng. Sci.}, 1995, \textbf{50},
  2669--2671\relax
\mciteBstWouldAddEndPuncttrue
\mciteSetBstMidEndSepPunct{\mcitedefaultmidpunct}
{\mcitedefaultendpunct}{\mcitedefaultseppunct}\relax
\EndOfBibitem
\bibitem[Kolb \emph{et~al.}(2004)Kolb, Cviklinski, Lanuza, Claudin, and
  Cl\'ement]{Kolb:2004}
E.~Kolb, J.~Cviklinski, J.~Lanuza, P.~Claudin and E.~Cl\'ement, \emph{Phys.
  Rev. E}, 2004, \textbf{69}, 031306\relax
\mciteBstWouldAddEndPuncttrue
\mciteSetBstMidEndSepPunct{\mcitedefaultmidpunct}
{\mcitedefaultendpunct}{\mcitedefaultseppunct}\relax
\EndOfBibitem
\bibitem[Zuriguel \emph{et~al.}(2005)Zuriguel, Garcimart{\'\i}n, Maza,
  Pugnaloni, and Pastor]{Zuriguel:2005}
I.~Zuriguel, A.~Garcimart{\'\i}n, D.~Maza, L.~A. Pugnaloni and J.~M. Pastor,
  \emph{Phys. Rev. E}, 2005, \textbf{71}, 051303\relax
\mciteBstWouldAddEndPuncttrue
\mciteSetBstMidEndSepPunct{\mcitedefaultmidpunct}
{\mcitedefaultendpunct}{\mcitedefaultseppunct}\relax
\EndOfBibitem
\bibitem[Candelier and Dauchot(2009)]{Candelier:2009}
R.~Candelier and O.~Dauchot, \emph{Phys. Rev. Lett.}, 2009, \textbf{103},
  128001\relax
\mciteBstWouldAddEndPuncttrue
\mciteSetBstMidEndSepPunct{\mcitedefaultmidpunct}
{\mcitedefaultendpunct}{\mcitedefaultseppunct}\relax
\EndOfBibitem
\bibitem[Varas \emph{et~al.}(2011)Varas, Vidal, and G\'{e}minard]{Varas:2011}
G.~Varas, V.~Vidal and J.-C. G\'{e}minard, \emph{Phys. Rev. E}, 2011,
  \textbf{83}, 011302\relax
\mciteBstWouldAddEndPuncttrue
\mciteSetBstMidEndSepPunct{\mcitedefaultmidpunct}
{\mcitedefaultendpunct}{\mcitedefaultseppunct}\relax
\EndOfBibitem
\bibitem[MacMinn \emph{et~al.}(2015)MacMinn, Dufresne, and
  Wettlaufer]{MacMinn:2015}
C.~W. MacMinn, E.~R. Dufresne and J.~S. Wettlaufer, \emph{Phys. Rev. X}, 2015,
  \textbf{5}, 011020\relax
\mciteBstWouldAddEndPuncttrue
\mciteSetBstMidEndSepPunct{\mcitedefaultmidpunct}
{\mcitedefaultendpunct}{\mcitedefaultseppunct}\relax
\EndOfBibitem
\bibitem[Rajchenbach(1990)]{Rajchenbach:1990}
J.~Rajchenbach, \emph{Phys. Rev. Lett.}, 1990, \textbf{65}, 2221--2224\relax
\mciteBstWouldAddEndPuncttrue
\mciteSetBstMidEndSepPunct{\mcitedefaultmidpunct}
{\mcitedefaultendpunct}{\mcitedefaultseppunct}\relax
\EndOfBibitem
\bibitem[Fischer \emph{et~al.}(2009)Fischer, Gondret, and Rabaud]{Fischer:2009}
R.~Fischer, P.~Gondret and M.~Rabaud, \emph{Phys. Rev. Lett.}, 2009,
  \textbf{103}, 128002\relax
\mciteBstWouldAddEndPuncttrue
\mciteSetBstMidEndSepPunct{\mcitedefaultmidpunct}
{\mcitedefaultendpunct}{\mcitedefaultseppunct}\relax
\EndOfBibitem
\bibitem[Yang \emph{et~al.}(2015)Yang, Li, Kong, Sun, Biggs, and
  Zivkovic]{Yang:2015}
H.~Yang, R.~Li, P.~Kong, Q.~C. Sun, M.~J. Biggs and V.~Zivkovic, \emph{Phys.
  Rev. E}, 2015, \textbf{91}, 042206\relax
\mciteBstWouldAddEndPuncttrue
\mciteSetBstMidEndSepPunct{\mcitedefaultmidpunct}
{\mcitedefaultendpunct}{\mcitedefaultseppunct}\relax
\EndOfBibitem
\bibitem[Nowak \emph{et~al.}(1998)Nowak, Knight, Ben-Naim, Jaeger, and
  Nagel]{Nowak:1998}
E.~R. Nowak, J.~B. Knight, E.~Ben-Naim, H.~M. Jaeger and S.~R. Nagel,
  \emph{Phys. Rev. E}, 1998, \textbf{57}, 1971--1982\relax
\mciteBstWouldAddEndPuncttrue
\mciteSetBstMidEndSepPunct{\mcitedefaultmidpunct}
{\mcitedefaultendpunct}{\mcitedefaultseppunct}\relax
\EndOfBibitem
\bibitem[Philippe and Bideau(2001)]{Philippe:2001}
P.~Philippe and D.~Bideau, \emph{Phys. Rev. E}, 2001, \textbf{63}, 051304\relax
\mciteBstWouldAddEndPuncttrue
\mciteSetBstMidEndSepPunct{\mcitedefaultmidpunct}
{\mcitedefaultendpunct}{\mcitedefaultseppunct}\relax
\EndOfBibitem
\bibitem[Bi and Chakraborty(2009)]{Bi:2009}
D.~Bi and B.~Chakraborty, \emph{Phil. Trans. R. Soc. A}, 2009, \textbf{367},
  5073--5090\relax
\mciteBstWouldAddEndPuncttrue
\mciteSetBstMidEndSepPunct{\mcitedefaultmidpunct}
{\mcitedefaultendpunct}{\mcitedefaultseppunct}\relax
\EndOfBibitem
\bibitem[Radjai \emph{et~al.}(1996)Radjai, Jean, Moreau, and Roux]{Radjai:1996}
F.~Radjai, M.~Jean, J.-J. Moreau and S.~Roux, \emph{Phys. Rev. Lett.}, 1996,
  \textbf{77}, 274--277\relax
\mciteBstWouldAddEndPuncttrue
\mciteSetBstMidEndSepPunct{\mcitedefaultmidpunct}
{\mcitedefaultendpunct}{\mcitedefaultseppunct}\relax
\EndOfBibitem
\bibitem[Silbert \emph{et~al.}(2002)Silbert, Erta\ifmmode~\mbox{\c{s}}\else
  \c{s}\fi{}, Grest, Halsey, and Levine]{Silbert:2002}
L.~E. Silbert, D.~Erta\ifmmode~\mbox{\c{s}}\else \c{s}\fi{}, G.~S. Grest, T.~C.
  Halsey and D.~Levine, \emph{Phys. Rev. E}, 2002, \textbf{65}, 031304\relax
\mciteBstWouldAddEndPuncttrue
\mciteSetBstMidEndSepPunct{\mcitedefaultmidpunct}
{\mcitedefaultendpunct}{\mcitedefaultseppunct}\relax
\EndOfBibitem
\bibitem[Mailman and Chakraborty(2012)]{Mailman:2012}
M.~Mailman and B.~Chakraborty, \emph{J. Stat. Mech.}, 2012, \textbf{2012},
  P05001\relax
\mciteBstWouldAddEndPuncttrue
\mciteSetBstMidEndSepPunct{\mcitedefaultmidpunct}
{\mcitedefaultendpunct}{\mcitedefaultseppunct}\relax
\EndOfBibitem
\end{mcitethebibliography}
\providecommand*{\mcitethebibliography}{\thebibliography}
\csname @ifundefined\endcsname{endmcitethebibliography}
{\let\endmcitethebibliography\endthebibliography}{}

}

\end{document}